%% file: 0-main.tex
\documentclass[nonacm, manuscript]{acmart}

\usepackage{hyperref}
\usepackage{graphicx}
\usepackage{color}
\usepackage{todonotes}
\usepackage[nolist]{acronym}
\usepackage{xcolor}
\usepackage{amsmath}
\usepackage{subcaption}
\usepackage[ruled,vlined,linesnumbered]{algorithm2e}

\usepackage{xcolor,colortbl}

\definecolor{Gray}{gray}{0.85}
\definecolor{LightCyan}{rgb}{0.88,1,1}

\newcolumntype{a}{>{\columncolor{Gray}}c}
\newcolumntype{b}{>{\columncolor{white}}c}

\SetKwProg{KwIn}{Input}{}{} 
\SetKwProg{Fn}{Function}{}{} 
{ \vspace{-0.15cm}%
    \small\noindent{\bfseries Availability of Data and Material:}\par%
    \noindent\ignorespaces}%
{ \par\noindent%
\ignorespacesafterend }%

\setcopyright{none}
\renewcommand\footnotetextcopyrightpermission[1]{}

\begin{document}

\title{From functioning to evolving: A complex systems perspective on future self-organised federated energy communities}

\author{Abdorasoul Ghasemi}
\email{abdorasoul.ghasemi@coventry.ac.uk}
\orcid{0000-0002-4432-4504}
\affiliation{%
  \institution{Coventry University}
  \city{Coventry}
  \country{U.K.}
}

\sloppy

\renewcommand{\shortauthors}{Ghasemi}

\begin{abstract}
Energy networks face a paradigm shift driven by renewable integration, uncertainty about required flexibility, distributed markets, and smart demand. Increasing asset interactions, cyber dependencies on communication and computation systems, and the deployment of AI agents demand a holistic, system-wide approach to understand the system's emergent behaviour. These transformations affect energy generation, demand adaptation, grid operations, and market dynamics, forming a complex engineered \emph{system of systems}. We argue that a complex-systems perspective can help to understand this evolution. Drawing lessons from two highly successful large-scale engineered systems, the Internet and agile software engineering, we highlight how prioritising design for evolution over traditional design for functionality enables energy systems to adapt to net-zero dynamics and handle unforeseen uncertainties. We discuss how the notion of federated energy communities (FEC), which aims for an open energy system with distributed coordination, aligns with this perspective. In FECs, each EC produces, stores, and manages its own energy, enabling it to trade with peers, respect grid constraints, and reach fair agreements. Beyond regulatory issues, the main challenge is identifying constraints that can orchestrate EC operations at the scale and reliability standards required of power networks, without unnecessarily limiting the innovation needed for system evolution. We then conclude with tentative design principles for future decentralised, self-organised energy networks through the lens of design for evolution, and how to apply them to a bottom-up architecture of autonomous energy communities.
\end{abstract}



\maketitle

\section{Introduction}\label{sec: introduction}
\input{1-introduction}

\section{System design: functioning or evolving}\label{sec: function-vs-evolving}
\input{2-functioning-evolving}

\section{Internet vs. PSTN for communications} \label{sec: Internet}
\input{3-Internet-Telephone}

\section{Agile vs. Waterfall methodology for software development}\label{sec: agile}
\input{4-Agile-Waterfall}

\section{Self-organized federated energy communities}\label{sec: self-organized-EC}
\input{5-distributed-FEC}

\bibliography{refs} 

\appendix

\end{document}

%% file: 1-introduction.tex
The \emph{Reductionist} hypothesis is the starting point for many analyses and designs of natural and human-made systems. It aims to reduce and explain the behaviour of every system by describing its components and their interactions, and has been successfully applied to construct complex engineered systems ranging from sophisticated microprocessor chips to aircraft. However, these great achievements sometimes lead to the illusion of \emph{constructionist}: ``The ability to reduce everything to simple fundamental laws does not imply the ability to start from those laws and reconstruct the universe'', because constructionism breaks when it is confronted with scale and complexity, ``More Is Different'' \cite{anderson1972more}. In complex systems, aggregate behaviour changes, and new behaviours may emerge as scale changes, suggesting the need to find new cause-and-effect rules.

Complexity manifests in different ways in complex systems, making it difficult to define a complex system. See \cite{bianconi2023complex} for a diverse perspective on the subject. Here, we adopt Newman's definition of a complex system as ``a system composed of many interacting parts, such that the collective behaviour of those parts together is more than the sum of their individual behaviours.''\cite{newman2011resource}.

The notion of complexity is mainly rooted in natural sciences and physics and is used to explain emerging behaviours in biological, social, and economic systems. However, the notion of complexity appears increasingly in the design and analysis of engineered systems like the Internet and large software systems. The traditional top-down reductionist/constructionist perspective and the adopted golden rule of \emph{separation of concerns} are not enough to explain some unpredictable behaviours in these systems, suggesting the need for a complementary complex-systems perspective \cite{mina2006complex}. 

The key point of the complex systems perspective on engineered system design is shifting the objective from \emph{functioning} to \emph{evolving} \cite{allen1988dynamic}. That is, the Newtonian mechanistic rationalism behind the traditional engineering design for \emph{closed systems} with fixed structures that obey a set of deterministic differential or partial equations is not valid for \emph{open systems} where new components may be added/removed, new interactions are created, and, more importantly, the system actively \emph{learns and adapts} itself. The system can adapt by learning latent patterns from large datasets using sophisticated artificial intelligence (AI) models, which system creators may overlook or fail to anticipate.    

Modern energy systems based on decentralised smart grids are changing rapidly for various reasons, including climate change policies, the massive integration of renewable sources and how they interact with the network, the rise of intelligent prosumers, the increasing demand for electric vehicles, and integration with other energy systems. These new requirements affect the long-term architecture of power systems, including large-scale generation, transmission, and distribution networks. In addition to integrating with ICT networks, advancing Internet of Things (IoT) technology and edge-cloud computing enable rapid adaptation to new requirements as a cyber-physical system. 

The transition to a modern \emph{open energy system of systems} with flexible pool energy management requires a new architecture and regulations that facilitate stakeholder interactions, including traditional fuel-based generators, transmission system operators (TSOs)/distribution system operators (DSOs), and new prosumers, ensuring the required flexibility for supply-demand balance. Nevertheless, these new interactions may lead to emerging collective and self-organised behaviours, such as instability in generation and markets, (de)synchronisation issues, cascading events/failures, and brittleness to exogenous disturbances. These behaviours suggest the need for a complex systems perspective to understand and characterise the system's behaviour profile.

\cite {witthaut2022collective} reviews emerging collective nonlinear dynamics and self-organisation in these highly penetrated renewable energy systems, providing a compelling justification for considering modern energy systems as complex systems. This work has a different objective: to identify design principles for the architecture of modern energy systems that facilitate \emph{evolution and responsiveness}. We first review how the complex system perspective (implicitly) supports a paradigm shift in analysing and designing the two most important human-made technological systems: the Internet and big software development methodologies. The transition from the public switched telephone network (PSTN) to the Internet as one of the most successful open networked systems and from the traditional waterfall methodology to agile methods as the current highly adopted software design method highlight how the complex systems perspective and, in particular, changing the objective from \emph{functioning} to \emph{evolving} is used in practice. These technologies involve massive interactions among heterogeneous system components and stakeholders, allowing system responsiveness to dominate its predecessors. Building on these observations, we then investigate how the principles of the Internet and Agile methodology, established by the founding fathers of these technologies, facilitate evolvability through the lens of complex systems. This paper aims to draw new insights for future energy systems architecture.

In the following, we first review and contrast the basic principles behind the classic reductionist/constructionist and complex systems perspective on system design in Section \ref{sec: function-vs-evolving}. In Sections \ref{sec: Internet} and \ref{sec: agile}, we examine how the complex systems perspective helps the transition from PSTN to the Internet and from waterfall to agile methodology. Section \ref{sec: self-organized-EC} builds on the lessons from \ref{sec: Internet} and \ref{sec: agile}  and discusses how the concept of distributed federated energy communities forwards the evolution of future energy systems. 

%% file: 2-functioning-evolving.tex
This section reviews some design principles, i.e.,  why the system is as it is, and outlines key characteristics of two perspectives on engineered system design: the classic reductionist/constructionist and the complex systems perspectives. Note that the objective is not to question the former approach, as almost all engineering achievements have benefited from it, but to emphasise that the second perspective could complement and provide new insights for system design, as we shall explain in our case studies in sections \ref{sec: Internet} and \ref{sec: agile}. Table \ref{table: two-views} summarises the conclusions of these two perspectives.
 
\begin{table}
 	\centering
    \caption{Constructionist and Complex systems perspective on engineered system design}\label{table: two-views}
 	{\renewcommand{\arraystretch}{2}
 		\begin{tabular}{ |>{\columncolor{blue!30}}p{5cm}|
 				>{\columncolor{yellow!30}}p{5cm}|
 				>{\columncolor{green!30}}p{5cm}|}
 			\hline
 			\textbf{} & \textbf{Constructionist view} & \textbf{Complex systems view} \\
 			\hline
 			Design principle & Functioning & Evolving \\ \hline
 			Goal & Optimality for designed-for uncertainties (highly engineered and robust)  & Responsiveness under unforeseen uncertainties (self-organized and resilient) \\ \hline
 			Methodology & top-down/separation of concerns (reductionist/constructionist)  & local interactions with facilitated protocols  \\ \hline
 			Assumptions  & closed systems, characterisable environment and uncertainty & open systems, responsive for unforeseen uncertainty   \\ \hline
 			
 			Process: 1)identifying the system requirements and constraints, 2)designing using the available and newly created tools, 3)testing and validation under planned conditions and environment, 4) deployment & Consecutive stages of (1)-(4)  & iterative process on steps 1 to 4 to evolve the solution, changing requirements over time \\
 			\hline
 			
 	\end{tabular}}
\end{table}

\subsection{Reductionist/constructionist system design}\label{subsec: reductionist-view}


The reductionist/constructionist perspective on designing engineered systems focuses on defining the system's objective. Here, it is implicitly assumed that the system is closed (no new components/interactions are added or removed), and that the environment and disturbances are predictable/determined. Hence, the system components/modules and their \emph{local} interactions are sufficient to infer the cause-effect rules and understand the system's behaviour at different scales. Practically, this viewpoint assumes that the big problem can be decomposed/reduced into smaller ones in a top-down approach, solving those small problems and then finding the whole solution by proper coordination/construction of those sub-problem solutions.

The underlying process for system development then consists of consecutive steps of 1) identifying the system requirements and constraints, 2) system design using the available and newly developed tools, 3) testing and validation under anticipated conditions, 4) manufacturing and deployment. The design goal is optimality: maximise (minimise) the system utility (cost) subject to constraints. The system utility (cost) encodes the desired objectives, and the constraints encode all physical, engineering (time/cost budget), and robustness to the designed disturbances. Therefore, the system is highly engineered to follow a single pre-determined optimal solution, and its performance is measured by how well it follows that solution across different scenarios.

Since the system structure and objective/constraints are fixed, system robustness should be planned at the design stage, which requires characterising the uncertainty: which system measures should be robust to which uncertainty. For engineering reasons, one should consider the worst-case disturbance and design the system to function in that scenario. Therefore, the achieved utility should typically be traded off against the considered uncertainty region: the larger the uncertainty region, the less desirable the achieved utility. 

The authors of \cite{carlson2000highly} argued that designing for high performance and robustness to designed-for uncertainty leads to a highly optimised tolerance mechanism and "robust yet fragile" characteristics: robust to foreseen uncertainty and fragile to unforeseen ones. 

Using the reductionist/constructionist design principle and relying on the separation of concerns rule, one can design standard interfaces/protocols between different parts of the system, which helps develop novel technologies for each part of the system without needing to account for the internal design and operation of the other system components, at the cost of missing a holistic perspective of the system operation and performance. \cite{srivastava2005cross} discusses the need for, and implementation challenges of, a holistic view of cross-layer protocol design in wireless networks to capture interactions across layers. More recently, \cite{chakraborty2021guest} discusses the holistic view in cyber-physical systems to capture tighter interaction between controller design and computer-based implementation. 

From the modelling perspective, uncertainties are modelled as independent probability distributions with limited variance, e.g., a Gaussian distribution, where the distribution's mean reflects the subsystem's expected behaviour and the effects of $n$ independent disturbances vanish with $1/\sqrt{n}$. This modelling implicitly implies bounded uncertainty in each subsystem and in the whole-system behaviour, which a proper strategy, such as engineered redundancies, could mitigate.  


\subsection{Complex systems perspective}\label{subsec: complex sys-view}

The system objective and design principle behind the complex systems perspective is \emph{evolvability}. Complex systems are often \emph{open}; new parts can be added and removed during their lifetimes, the environment is partially predictable, and they may encounter unforeseen disturbances. Therefore, the system should \emph{adapt on a short time scale} and \emph{evolve on a long time scale} to manage unforeseen disturbances and new requirements.

The system's design goal is responsiveness; it should respond properly even to unforeseen disturbances. The design process is evolutionary and iterative, following the four system-design steps mentioned in the previous section. No single solution exists for the big problem in a complex system, as the system's objective and constraints may change over time. Therefore, the structure (bottom-up architecture) and function (top-down organisation) of complex systems co-evolve. The system is self-organised and manifests self-x features such as self-healing and self-recovery. Self-organised order emerges from limited, simple interactions, system constraints, and protocols (regulatory processes) that push the system toward a solution. 

A salient feature of evolving systems is their ability to learn and adapt. Evolving systems can adapt after deployment and replace the traditional design-and-deploy approach with design-deploy-and-adapt to cope with diverse uncertainties and respond properly \cite{eiben2015evolutionary}, materialising the difference between ``functioning'' and ``evolving'' \cite{allen1988dynamic}. 

The architecture favours adaptation over optimality, as natural evolution selects species that can cope with change rather than those with optimal behaviour \cite{allen1988dynamic}. The need for adaptation and restructuring, inserting and removing new components/interactions, is manifested in the system structure, such as hierarchical, sparse, and favouring special subgraphs (motifs).   

Protocols or regulatory processes facilitate the bottom-up organisation of these systems. Protocols define the rules for interaction among system components, imposing additional constraints and reducing the possible number of solutions \cite{alderson2010contrasting}. That is, beyond component- and system-level constraints, a good set of protocols \emph{minimally constrains} the solution space and helps the system respond quickly by finding a feasible, high-quality solution. The Internet architecture mainly relies on two main protocols:  Internet Protocol (IP) for scalable end-to-end packet delivery (reachability) and the Transmission Control Protocol (TCP) for end-to-end process communications (reliability), as we will discuss in Section ~\ref{sec: Internet}, allowing effective communication among billions of end-users with diverse behaviours and uncertainties. The hourglass shape of the Internet protocol stack captures the idea of protocols as ``constraints that deconstrain'' \cite{gerhart2007theory}, facilitating the search for solutions under different conditions by properly reducing the solution space. 

Many behaviours in complex systems are modelled by power-law distributions, which are observed as a universal signature in the structure or dynamics of complex systems. Using power-law distributions with infinite variance, or even mean, implies unbounded uncertainty from a modelling perspective. That is, the risk of rare events in systems with power-law distributions is high \cite{sornette2009probability}. Complex systems are inherently unpredictable, and we should prepare for unforeseen events. The prominent example of such a high-risk rare event is cascading failure in power networks. This unpredictability requires preparedness that may waste resources in the short term but pays off in the long term. 

Therefore, unlike functioning systems, complex systems maintain responsiveness not only through robustness and proper inclusion of redundancies, but also through other aspects of resiliency such as adaptability, extensibility, and rebound \cite{woods2015four}. Learning and adaptation are well-known evolutionary strategies that need to be implemented in complex engineered systems.










\subsection{The need for complex systems perspective}
This section explains why energy networks, as engineered systems, should be considered a complex system of systems and how complexity science can help us better understand their evolution.

We briefly list some reasons, which are not exhaustive and may overlap/correlations. The argument is that emerging power networks may face disturbances that require greater complexity/degree of freedom to respond to increasingly complex scenarios, according to Ashby’s law of requisite variety \cite{ashby1991requisite}. 

This law provides a necessary but not sufficient condition for a system to respond properly to its environment. According to the law of requisite variety, the system's number of states (i.e., complexity) must equal or exceed the environment's number of states. That is, the system must have a degree of freedom for each possible environment behaviour. 
The multi-scale version of Ashby's law of requisite variety states that system complexity must equal or exceed that of its environment at all scales \cite{siegenfeld2025formal}. 

As we discuss later in Section~\ref{sec: self-organized-EC}, self-organised energy communities could potentially provide this freedom.

\begin{itemize}
\item Interactions with social and economic systems: Energy systems will be affected by complex social and economic systems in the coming years. For example, how will the many new renewable prosumers self-organise to respond to new regulatory processes or social and economic disturbances, thereby affecting the behaviour and even the stability of the power grid? If a single provider, or a small group of solar energy providers, does not sell its energy, it will not significantly affect the system's operation. However, if many of them coordinate on social media and make collaborative decisions, they could affect the whole system and its stability. Here, complexity science helps predict complex social and economic patterns and their impact on energy systems.

\item Interaction with other technological systems, such as communication and distributed digital control (cyber interdependency): energy systems are interdependent systems in which new behaviours may emerge from nonlinear interactions between systems. In particular, the behaviour of failure cascading, one of the most important risks in power networks, changes with interactions, and these interactions can amplify how failures/malfunctions propagate across networks. The notion of complex interdependency among critical infrastructures and the importance of energy systems as the heart of the interdependent critical infrastructure ecosystem, which directly and indirectly influences/gets influenced by other infrastructure systems, has been known for decades \cite{rinaldi2001identifying}. Interdependencies create subtle interactions and feedback; uncontrolled interdependencies pose a systemic risk to system resiliency. From a network-connectivity perspective, interdependent networks are more fragile \cite{buldyrev2010catastrophic}. Also, considering the power and communication flows over networks, the type (strong or weak) and degree of coupling between the power and communication networks change the behaviour of the interdependent power-communication system \cite{ghasemi2023robustness}	

\item Unpredictable/irreducible uncertainties: The scope (due to interdependencies) and nature (due to increased penetration of renewable generation) of uncertainties in energy systems increase. For example, weather predictability is limited; therefore, the amount of available wind and solar energy is not known exactly or even predictable for a 10-day-ahead long-term forecast \cite{krishnamurthy2019predictability}. 

Combined with other technological, social, and economic interactions, it is not possible to characterise or enumerate all possible scenarios the system may encounter, suggesting that the system's responsiveness is critical. Data-driven approaches and AI can definitely help to reduce these uncertainties. However, the system should be responsive to a certain amount of irreducible uncertainty.



		
\item Deployment of AI and Human-AI interaction: Energy systems increasingly deploy micro-enabled decision-making agents. AI helps find and use solutions/patterns in energy consumption/generation/prediction and user behaviours that systems' creators overlook or cannot distinguish. However, the collective behaviour of these decisions is unclear across the whole system. Furthermore, the control room is expected to rely more on AI recommendations/decisions, which a human with limited cognitive capacity should approve, meaning humans' and machines' decision-making uncertainties are part of the system. 
\end{itemize}

%% file: 3-Internet-Telephone.tex
The Internet is one of the most successful man-made systems, and its complexity is comparable to natural systems. Internet started as a small project with fixed endpoints, eventually connected billions of endpoints, and became a pervasive, transformative technology that affects all aspects of human life, including social, economic, and technological aspects.

The Internet's structure and function are evolving to meet new requirements and expectations, and it is also changing our expectations of communication anywhere, anytime, with anyone. The best way to understand the Internet's evolution is to compare it with its predecessor, the public switched telephone network (PSTN), which has functioned well for many decades with little change. This comparative study may also offer insights into the transition from the current highly engineered generation/transmission/distribution energy network structure to modern architectures suitable for the green transition. 

In the following, we compare the Internet's design principles, established by the Internet's founding fathers, with those of the PSTN, and interpret them through the lens of complex systems concepts to draw insights for engineered complex systems. Table~\ref{table:pstn-internet} summarises the main differences from a systems-thinking perspective.

\subsection{From PSTN to the Internet}
The PSTN has a clearly hierarchical architecture, and its function depends on two separate consecutive stages: design and deployment. Design relies on accurate traffic-matrix forecasting to provision sufficient resources and allocate them properly during deployment across different network levels. Resource allocation to calls is completely network-centric (intelligent network), relying on standard signalling between exchanges, while the dumb end-points (telephones) are passive components. The single objective is minimising blocking probability. PSTN services have largely stayed the same for decades, and PSTN uncertainties are limited mainly to device failure. PSTN architecture relies on admission control and an instantaneous demand-supply balance without buffers. The selected routes are entirely predictable by predefining the main path and the alternatives. 

The Internet, however, has a loose architecture that mainly relies on the highly scalable simple Internet Protocol (IP) as a focal point and exploits the available infrastructure of the time-pervasive PSTN. The Internet evolves over time; resource planning relies on over-provisioning, and resource allocation to flows is mainly user-centric, with possible network impacts (intelligent user/semi-dumb network) for traffic shaping and policing. Let's say edge-centric.

The endpoint Transport Control Protocol (TCP) handles congestion control based on implicit loss/delay feedback from the network. (Remember the concept of Grid-friendly Energy communities). Although IP has changed only marginally (except for the migration from IPv4 to IPv6), new TCP protocols are introduced as the Internet evolves, improving resource utilisation. 

The objective of the whole network is reachability between every two endpoints: IP ensures connectivity across the system, and TCP adapts endpoints' resource share from the network (their rates) to avoid congestion and resource under-utilisation. As far as an endpoint can send/receive IP packets and the corresponding application can effectively communicate through TCP or user datagram protocol (UDP- an unreliable transport protocol that simply demultiplexes different applications), the other details on the data link/physical layer protocols and application protocol do not matter and are up to the local network and end-applications. 

IP separates the constantly evolving physical technology at the lower layers and new applications and requirements at the upper layers. The emerging self-organised behaviours of highly engineered yet complex systems, such as the self-similarity of burst patterns in traffic and the scale-free nature of the degree distribution of autonomous systems (ASs) structure, then appeared as the result of complex interactions of the end-users, components, and protocols \cite{willinger2002scaling, alderson2005contrasting}.


In contrast to PSTN, node buffer size and buffer management are crucial for resource management on the Internet. The Internet adapts more quickly by shifting adaptations to endpoints and the network edge, mostly through software updates; this enables exploiting the crowd creativity at the endpoints to cope with emerging requirements and provide new services. Note that this adaptation is possible because the network core operation based on IP is simple, since any change in the network core requires a huge amount of investment for updating the hardware and takes time to be adopted by important players.

At the largest scale, the Internet is an interconnection of Autonomous Systems (AS)  that communicate using the Border Gateway Protocol (BGP) under provider-customer or peer-to-peer relationships, where each AS sets its own internal policy independently. BGP is highly scalable, and despite each AS's independent policies, the system operates reliably. Each AS consists of small- and large-scale Internet Service Providers that arrange their mutual contracts independently and also serve end users.  

The sources of uncertainty in the Internet are enormous across different time and spatial scales, ranging from component failure and intentional (distributed) denial-of-service attacks to ASs policies. For example, in a seminal paper, Paxson analysed the large-scale behaviour of end-to-end routing in the Internet at the end of 1995, including routing stability and the likelihood of encountering a major routing pathology, concluding that Internet routing has become less predictable in major ways compared to the end of 1994 \cite{paxson1996end}. However, the Internet adapts over time thanks to its evolvable architecture.    
\begin{figure}
	\centering
	\includegraphics[width=1.1\textwidth]{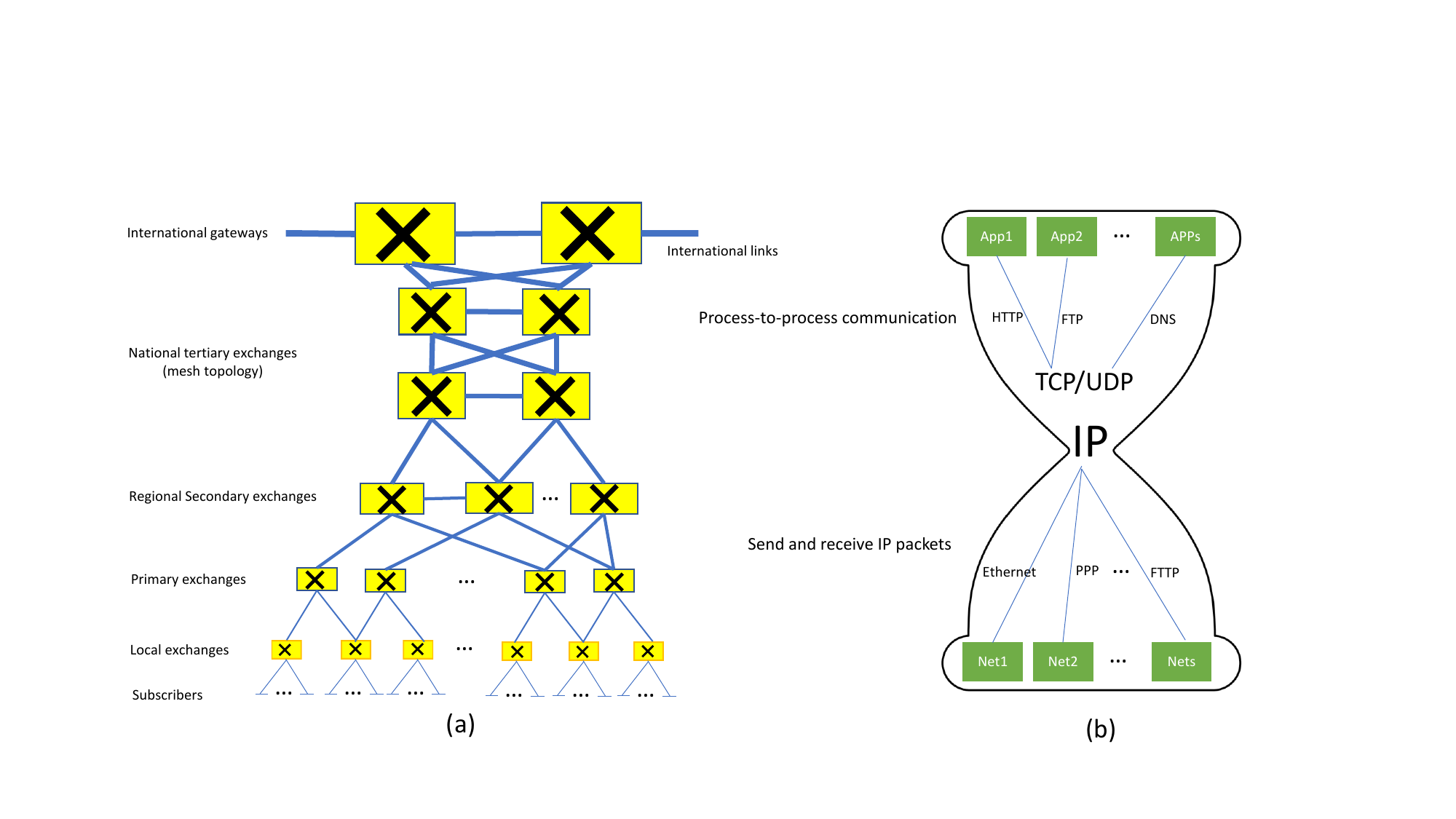}
	\vspace{-15mm}\caption{The architecture of (a) hierarchical PSTN and (b) hourglass Internet}
	\label{fig: pstn-internet}
\end{figure}

\subsection{Internet from complex system perspective}
The Internet's successful evolution is rooted in the insights of its early architects, who shaped its design philosophy and raised the question of whether its design principles can be generalised to other open networked systems, such as modern energy systems.

The Internet's key design principle is defining the priority of system goals. The first-level goal is to ``develop an effective technique for multiplexed utilisation of existing interconnected networks'' \cite{clark1988design}, rather than the tempting objective of optimising the utilisation of different transmission media. This choice led to adopting packet-switching technology for interconnecting distributed, independently administered networks. 

The next three second-level goals, which have a profound impact on Internet architecture, are in order: 1) survivability in the face of failure, 2) offering multiple types of service, and 3) accommodating a variety of networks. If the order and importance of these goals were different, the Internet would have a different architecture. These goals were eventually encompassed in the hourglass-shaped TCP/IP architecture. Interestingly, optimal resource (at the time very expensive) utilisation and accountability were not top priorities, but were eventually achieved by introducing the ``soft state'' concept for flows \cite{clark1988design} and by supporting different service models beyond the initial best-effort service model \cite{shenker1995fundamental}.  

Therefore, the Internet's design principles and goals align with the complex systems perspective. 
The Internet is also an open system that evolved iteratively as new requirements and constraints emerged, favouring simplicity that enables adaptability over optimality and materialising a complex system \cite{PetersonDavie2021}. 

In \cite{alderson2010contrasting}, the authors searched for universal laws and examined the role of system architecture in highly evolving, technologically complex networked systems. The authors consider architecture as a means for handling component-, system-, and possible emergent-level constraints or imposing constraints by protocols, to help system evolution in finding a solution compatible with its goals. Here, the emphasis is again on the impact of protocols (TCP/IP for the Internet) as ``constraints that constrain'', which confine the solution space in a way that facilitates variation and robustly finding a feasible solution. 





%
%

\begin{table}
	\centering
	\caption{PSTN functioning vs. Internet evolving as two engineered systems}
	\label{table:pstn-internet}
	{\renewcommand{\arraystretch}{2}
		\begin{tabular}{ |>{\columncolor{blue!30}}p{5cm}|
				>{\columncolor{yellow!30}}p{5cm}|
				>{\columncolor{green!30}}p{5cm}|}
			\hline
			\textbf{} & \textbf{PSTN (functioning)} & \textbf{Internet (evolving)} \\
			\hline
			Architecture & Detailed hierarchical & Loosely protocol-based with a focal TCP/IP protocol \\ \hline
			Objective & Optimality: minimise probability of blocking given available resources & Responsiveness: reachability, find a path between source and destination \\ \hline
			
			Design methodology and process & top-down, consecutive stages for functioning systems  & local interactions are coordinated by protocols, iterative process for evolving systems  \\ \hline
			
			Resource supply demand balance & accurate traffic forecasting at design stage (long-term) and call admission control, no buffer (short-term) & rough provisioning at design stage (long-term) and traffic control at end-point and shaping/policing at edge- buffer management is crucial (short-term)\\ \hline
			Resource allocation  & network centric (intelligent network-dumb end points) & user(edge)-centric (intelligent end-points-dumb network) \\ \hline
			
			Design process and methodology & top-down and consecutive & iterative and bottom-up \\
			\hline
	\end{tabular}}
\end{table}

%% file: 4-Agile-Waterfall.tex
Another prominent example of engineered complex systems is big software with hundreds or thousands of interacting modules and interfaces. There was a paradigm shift 
from traditional \emph{waterfall} to \emph{agile} methodology for software development. Agile methodology transforms the software development industry by shifting the goal from \textit{optimisation} to \textit{adaptation} \cite{nerur2007theoretical}. 

The design philosophy of agile software methodology is anchored in the ``Agile Manifesto'' \cite{beck2001manifesto}, which prioritises individuals and interactions over processes, working software over documentation, customer collaboration over contract negotiation, and responding to change over following a rigid plan.

\subsection{From waterfall to agile}
The classic software or waterfall methodology was used for decades, grounded in technical/functional rationality and logical positivism that assumed a stable, predictable environment.

Under this mechanistic paradigm, waterfall methodology treats software development as a controllable, predictable process managed through upfront planning that must strictly follow systematic procedures. Therefore, 1) the primary objective is optimality in selecting the best tools to achieve the planned activities, and 2) development is organised as a deliberate, formal, linear sequence of steps in which upfront planning and control can handle uncertainties \cite{dyba2009we, nerur2007theoretical,kuhrmann2021makes}.

Waterfall methodology follows the classic "separation of concerns" engineering principle by separating design from implementation, requiring detailed design before implementation. This methodology assumes the system requirements, architectures, and project risks can be completely identified, analysed, and specified before the implementation phase. Development follows a top-down approach in which the project manager acts as a controller who directs tasks and procedures, prioritises control, and relies on interchangeable standard roles.      

Agile, however, treats design and implementation as inseparable phases that evolve together, where solutions cannot be fully planned in advance. Agile is a dynamic human-centred methodological framework that prioritises adaptation and responsiveness over optimisation \cite{dyba2009we}\cite{nerur2007theoretical}. It replaces top-down control by positioning managers as facilitators rather than controllers and enabling self-organising, heterarchical team development and active stakeholder engagement. 

Agile relies on iterative, incremental software development rather than waterfall, plan-driven approaches. It treats design and implementation as inseparable activities that evolve iteratively together. Agile approaches problem-solving through continuous learning, experimentation, and introspection while constantly reframing both the problem and its solution. The Agile framework assumes a turbulent, highly complex, and difficult-to-predict operating environment and handles unpredictability. 

\subsection{Agile from a complex system perspective}

Agile frameworks question the assumption that a high degree of formalisation can control change and uncertainty. In fact, according to Ashby’s law of requisite variety, such formalism limits the degree of freedom the software development team needs to respond to the environment.  

Agile views software development through a complex systems lens, replacing deterministic control with dynamic, emergent, and self-organising paradigms. As a complex system, software development results from the co-evolution of problem definition and solutions \cite{nerur2007theoretical}.   

Agile relies on decentralised, autonomous, self-organising units that allow local adaptation to environmental disturbances, generating distinct responses. This autonomous, self-organised behaviour amplifies system variety and thus helps it respond to external disturbances \cite{nerur2007theoretical}.

\begin{table}
	\centering
    \caption{Waterfall vs. Agile methodology as two software development methodologies}

	{\renewcommand{\arraystretch}{2}
		\begin{tabular}{ |>{\columncolor{blue!30}}p{5cm}|
				>{\columncolor{yellow!30}}p{5cm}|
				>{\columncolor{green!30}}p{5cm}|}
			\hline
			\textbf{} & \textbf{Waterfall (functioning)} & \textbf{Agile (evolving)} \\
			\hline
			Architecture & degree of formalisation can control/detailed centralised hierarchy,  &  decentralised, autonomous, self-organising\\ \hline
			Goal & Optimality (e.g., minimise project risk/required resources)  & Responsiveness/adaptation, flexibility \\ 
           
			\hline
			Design methodology & Top-down functional decomposition; systematic and standardised procedures;  formalised innovation;  design precedes code  & continuous reframing of problems and solutions;  collaborative, dialectic, opportunistic exploration; manager acts as facilitator; design and code evolve concurrently \\              \hline
            Design process & Formal specification and linear; rule-driven with separate implementation phases & Emergent, iterative, and exploratory; action and knowing are inseparable \\\hline
			Assumptions  & stable and predictable environment & turbulent, difficult-to-predict environment \\ \hline

	\end{tabular}}
	
\end{table}\label{table: waterfall-agile}

%% file: 5-distributed-FEC.tex


This section explains how the EU Energy communities (ECs) notion \cite{ec_energy_communities} and federated ECs can facilitate the shift toward a bottom-up, self-organised system with the variety needed to respond effectively to different disturbances.  

Energy communities (ECs) \cite{ec_energy_communities} represent a major shift in how renewable energy generation is organised and governed in the interest of owners. Instead of relying on a centralised point to decide how much energy is generated and when, local communities manage their own renewable energy production, storage, trading, and distribution. These decentralised ECs aim to be as self-sufficient as possible while still interacting with the wider electricity grid and with other communities.

The UK and European policies are actively promoting ECs as part of the net-zero transition. However, scalable coordination models and mechanisms remain missing. 

The main challenge is how many autonomous communities can coordinate fairly and efficiently without creating grid congestion or instability. Current systems are not designed for this bottom-up transformation. Without new coordination mechanisms, the rapid growth of local renewables could strain existing infrastructure. What main constraints should ECs follow to ensure the reliability expected of the energy industry? 

\subsection{From Legacy Grids to Evolving Networks}

From an architectural perspective, like the PSTN, traditional power networks rely on dumb consumer endpoints equipped with a meter and highly engineered systems of power generation, transmission, and distribution, without the capability of power storage, and therefore require instantaneous power generation–demand balance.

Modern energy systems should instead consider distributed smart prosumers (producer/consumer) with distributed, intermittent renewable energy sources. Furthermore, electric power storage technology is advancing, enabling better response to peak/burst demand or unexpected generation shortfalls. In parallel, the increasing deployment of distributed energy storage adds flexibility and controllable resources while also demanding new coordination and operational requirements across the network.

From an operation-and-control perspective, like the waterfall framework, the traditional operational perspective relies on centralised and strict monitoring/control rules and detailed, predetermined responses. However, the number of potential environmental disturbances increases daily as more inverter-based renewable resources and AI-based loads are added to the network. 

This raises the question of whether centralised control can respond efficiently at scale, or whether it could impede the innovation that could otherwise emerge in evolving distributed systems. 

This paper argues that transforming the network and its operation to accommodate these changes requires a shift from a centralised, top-down system to a bottom-up, decentralised, evolving system. This transition requires defining priority system goals at different levels, as well as the architecture and protocols, to support the long-term evolution of the whole system.

Different studies investigate power system evolution scenarios and enabling technologies across regions. The authors of \cite{knezovic2021power} studied the long-term evolution of power systems by identifying the macro trends behind energy transition and corresponding challenges.

From a complex-systems perspective, instead of trying to list/find/forecast the possible evolution scenarios, we should facilitate evolution by properly designing protocols among stakeholders, including prosumers, DSOs, TSOs, Traditional/Renewable generations, and the regulator. 

\subsection{Federated ECs}
Federated ECs (FECs) suggest a new framework for self-organised ECs. Rather than relying on a central authority, autonomous communities will coordinate among themselves using a distributed consensus mechanism, enabling them to share energy, respect grid constraints, and reach fair agreements. 

Transition to EC architecture, however, is challenging, as it must address deep uncertainties ranging from the availability of renewable energy sources to smart demand and a fluctuating energy market in a distributed manner. 

The self-organised FECs must 1) ensure the required reliability level for the power industry, 2) be market-aware and cost-effective, and 3) respect the grid constraints and avoid violations. 

\begin{figure}
	\centering
	\includegraphics[width=1.0\textwidth]{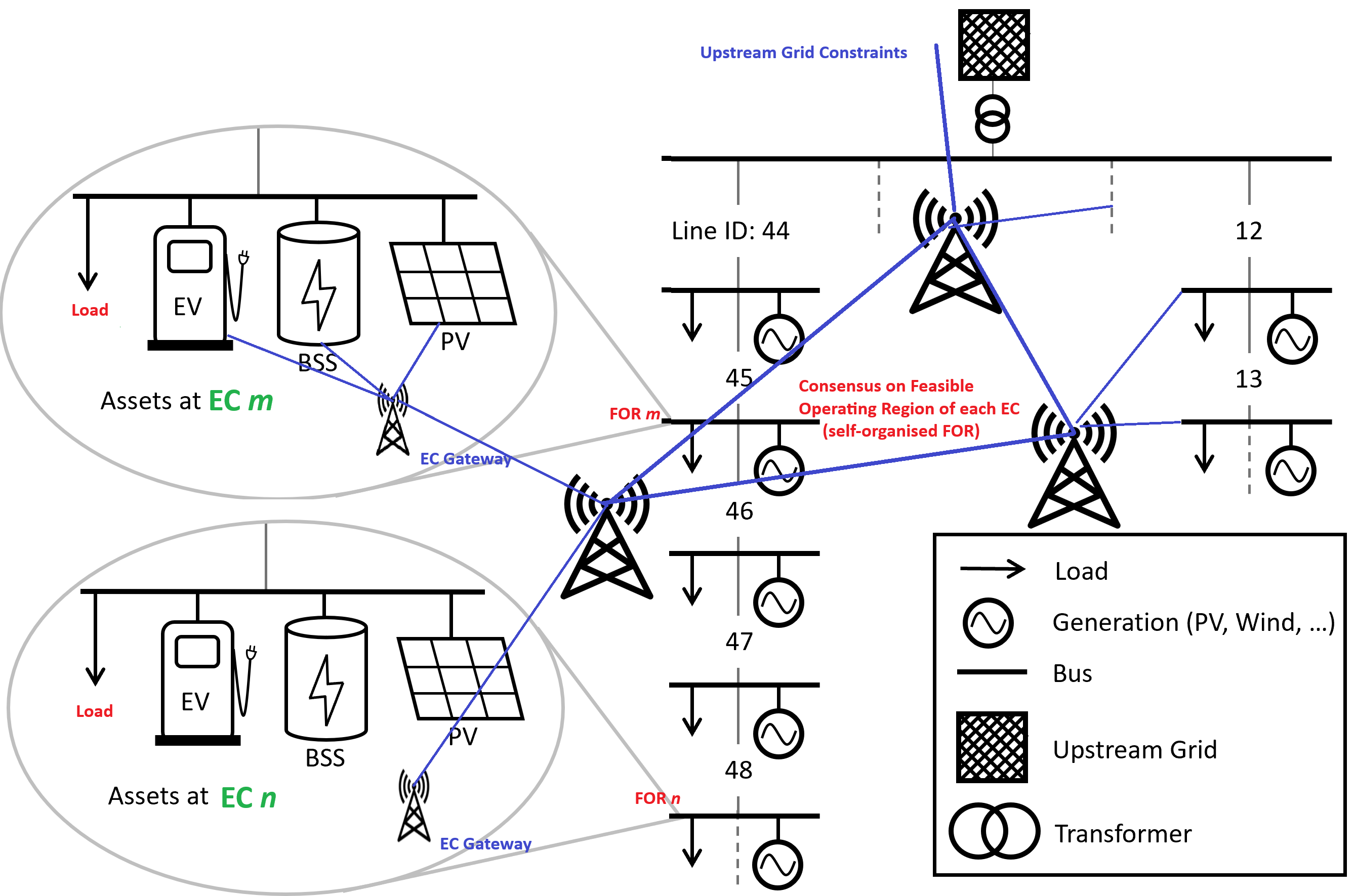}
	\caption{The schematic diagram of a self-organised federated energy community. Each EC must adhere to the advertised Feasible Operating Region (FOR), determined by the ECs' advertised demand/generation and uncertainty level, network congestion constraints, and upstream grid constraints.}
	\label{fig: self-organised energy Communities}
\end{figure}

In FECs, each autonomous EC has a speaker agent that communicates messages to reach consensus on binding constraints while respecting upper-layer grid limits.
See Fig. \ref{fig: self-organised energy Communities}.

The federated speakers will reach a consensus on the binding feasible operating region (FOR) for each EC using a self-organised distributed algorithm. The FOR can, for example, identify the active/reactive power injection, ramping limits, or voltage tolerance. The consensus process is iterative: ECs submit their offers/constraints to their speakers, and after consensus, the speaker returns the agreed FOR to the ECs. The consensus decision accounts for hard upstream grid constraints, soft grid congestion constraints, the current energy market, and the price of using the grid to encourage self-consumption and promote self-sufficiency.

After the convergence and determination of each EC's FOR, the internal decision on how to adhere to it is entirely up to the EC: asset owners in that EC can decide internally how to balance their interests regarding energy demand time shifting, trading preferences, and battery management systems. They are also autonomous in meeting the agreed FOR: how to micro-manage and enforce each household or asset manager's compliance with the FOR using customised, agreed-upon fairness criteria. 

The FEC draws on different approaches to respond to renewable energy curtailment while adhering to network constraints and market awareness in a scalable solution. Primarily, it draws on principles of scalable architecture design for complex networked systems \cite{alderson2010contrasting}. Technically, it stands on previous studies on EC operation in interaction with the grid \cite{sudhoff2024operation, lechl2026twostage} and peer-to-peer energy trading \cite{morstyn2018using}.

The authors of \cite{sudhoff2024operation} focused on the grid-interactive use of renewable community participants’ flexibility to support the distribution system. The ECs offer DSOs the flexibility they have available, and DSOs can accept or request adjustments after solving a centralised optimal power flow problem. In \cite{lechl2026twostage}, the authors show how to control a single EC  in a two-stage process: the global controller derives constraints that ensure congestion-relieving behaviour, while a local controller at each EC optimises resource setpoints to maximise self-consumption and revenue.

FECs will also use the bottom-up solution from the desired peer-to-peer energy trading literature \cite{morstyn2018using}, which allows individual prosumers to decide about their own generated energy/demand. P2P energy trading, however, is not scalable and lacks liability and responsibility in the event of non-delivery because of the complexity of imposing a regulatory framework. The FEC approach is a decentralised, realistic solution that relies on edge continuum computing as the new computing paradigm for smart grids.

\begin{table}
	\centering
    \caption{Traditional and Energy Community-based power networks: functioning vs. evolving as two engineered systems}
	{\renewcommand{\arraystretch}{2}
		\begin{tabular}{ |>{\columncolor{blue!30}}p{5cm}|
				>{\columncolor{yellow!30}}p{5cm}|
				>{\columncolor{green!30}}p{5cm}|}
			\hline
			\textbf{} & \textbf{Traditional  networks (functioning)} & \textbf{FECs networks(evolving)} \\
			\hline
			Architecture & Detailed centralised hierarchal  generation-transmission-distribution  & distributed self-organised ECs \\ \hline
			Objective & Optimality: minimise the generation cost & Responsiveness: meet the demand \\ \hline
			
			Design methodology and process & top-down, strictly functioning systems  & local interactions among ECs and DSOs, iterative process for evolving systems  \\ \hline
			
			Resource supply demand balance & accurate centralised and preplanned at different time scales & relies on self-sufficiency of distributed ECs and peer-to-peer energy trading among ECs and DSOs \\ \hline
			Resource allocation  & network centric (intelligent network-dumb end points) & EC-centric (intelligent ECs and facilitating network) \\ \hline
					
			\hline
	\end{tabular}}
\end{table}\label{table: pstn-internet}


\subsection{FEC as evolving systems}

The notion of ECs aligns with the complex systems perspective on energy networks. An EC-based architecture represents a paradigm shift in energy systems, restructuring the whole system as a bottom-up, self-organised system \cite{ec_energy_communities}. ECs will address two major interrelated challenges in the transition to net-zero: ensuring the demand-supply balance with distributed, intermittent renewable energy sources and upgrading the network to accommodate power transfer. These challenges stem from the power system's top-down architecture and operation, based on large power plants and their components.  In this paradigm, evolving energy systems will consist of distributed energy communities that are as self-sufficient as possible and grid-friendly, avoiding grid congestion.

The whole system consists of interconnected ECs that are as self-sufficient as possible: the distribution grid should then facilitate this interconnection by installing local battery systems in good locations to accommodate the required predicted flexibility. 


